# A Chain-of-Thought Prompt Engineering Framework for Automated Technical Specification Generation in Enterprise Integration: A Real-World Case Study

**Aryan Dhingra[1], Gokul Menon[1], and Anusha Antony[1]**

[1]Khoj Information Technology, Ramsey, NJ 07446 USA
Corresponding author: Aryan Dhingra (e-mail: aryan.dhingra@khoj-inc.com).

This work was carried out as part of normal commercial delivery activity; it received no external research funding.

**ABSTRACT** Technical Specification Documents (TSDs) translate functional requirements into implementable designs and are mandatory artifacts in governed Enterprise Application Integration (EAI) programs, yet their production remains a manual, expertise-intensive bottleneck. Large Language Models (LLMs) have been studied extensively for code generation, but long-form, template-governed engineering documents remain comparatively unexplored. This paper reports an industrial case study, conducted during a migration of more than 250 enterprise integrations to Microsoft Azure Integration Services at a large Australian wholesale distributor, that introduces a chain-of-thought prompt-engineering framework for automated TSD generation using GitHub Copilot. The framework combines a two-phase generation architecture that grounds the model in the functional specification, organizational standards, and the approved template; a six-stage prompting pipeline with human validation checkpoints; and Prompt-as-a-Code, under which prompts are version-controlled and reviewed as engineering artifacts. Applied to a stratified sample of fifteen production interfaces, the framework reduced TSD authoring effort from 793 to 300 person-hours, a 62.2% pooled reduction and a 64.0% mean per-document reduction (Wilcoxon signed-rank, $n = 15$, $p < 0.001$), at 74.3% reviewer-assessed accuracy and 3.7/5 developer satisfaction. Benefits degrade monotonically with interface complexity: between the simplest and most complex bands, effort reduction falls from 70.6% to 58.9% and accuracy from 91% to 62.5%, and the residual human effort concentrates in diagrams, sequence flows, and cross-system integration content. We additionally report comparative observations of GPT-4.1/5 and Claude Sonnet 4, as well as Copilot's Ask and Agent interaction modes, offering reusable, practice-grounded guidance for enterprise settings.

## I. INTRODUCTION

Enterprise integration programs are document-driven. Before an interface is built, a Technical Specification Document (TSD) must translate the business-facing Functional Specification Document (FSD) into an implementable design: processes and sequence flows, component architecture, data mappings, exception handling, security controls, observability, deployment, and disaster recovery. In governed delivery organizations the TSD is a gating artifact; it is reviewed against architecture principles and an approval checklist before development may proceed. When an organization must deliver integrations at scale, this documentation discipline becomes a substantial cost and time drain. The setting for this study is exactly such a program: an enterprise-wide modernization at a large Australian wholesale distributor, in which more than 250 integrations are being rebuilt or migrated from a legacy TIBCO platform to Microsoft Azure Integration Services (AIS), including 117 newly built supply-chain interfaces, roughly 106 migrated retail interfaces, and over 60 finance interfaces. In this program, the two interfaces on which the framework was first developed carried a combined manual TSD-authoring baseline of 74 person-hours, effort that recurs, with variation, for every interface in the portfolio.

Large language models (LLMs) have transformed expectations for software-engineering productivity, but the research emphasis has been on code: completion and synthesis [1], developer productivity with AI pair programmers [2], [3], test generation, and program comprehension [4]. Long-form engineering and technical design documents occupy a different regime. A TSD is not free-form prose; it must mirror a mandated template, comply with organizational standards and naming conventions, remain aligned to a source of FSD, embed diagrams and structured tables, and withstand formal review. Naive single-pass prompting of an LLM against such requirements tends to produce plausible but generic text, skipped sections, and hallucinated design details. Whether disciplined prompt engineering, in particular chain-of-thought (CoT) decomposition [5], can close this gap in a real delivery organization is an open, practice-relevant question on which industrial evidence is scarce.

This paper reports an industrial case study of a CoT prompt-engineering framework for TSD generation, designed and applied inside the distributor's AIS delivery organization during a structured GitHub Copilot adoption program (July–October 2025, with demonstrations through January 2026). The framework was developed on two production interfaces, identified by their Enterprise Application Integration Interface (EAII) catalogue numbers: EAII-270 (“Client Identity”, an Azure Function App) and EAII-265 (“Employee Info Feed”, an Azure Logic App integration), whose outputs were adopted as the delivery artifacts of record after human review, and was subsequently applied across a stratified sample of fifteen. We investigate three research questions:

**RQ1.** Can chain-of-thought prompt engineering with a general-purpose AI coding assistant reliably generate standards-compliant TSDs from FSDs in a production enterprise setting?

**RQ2.** How do different foundation models (GPT-4.1/5 versus Claude Sonnet 4) and interaction modes (Copilot Ask versus Agent mode) compare for this task?

**RQ3.** What efficiency, accuracy, and satisfaction outcomes are achievable in production, how do they vary with interface complexity, and which parts of the document still require human effort?

The study makes four contributions. First, it specifies a reusable framework for LLM-based generation of template-governed enterprise documents: a two-phase architecture (basic generation of ~75% of the document, then advanced enrichment of the remaining ~25%), a six-stage CoT pipeline with human validation checkpoints, and Prompt-as-a-Code (PrAAC), which places prompts under version control and code review in Azure DevOps. Second, it reports production outcomes across fifteen documents: a 62.2% pooled reduction in TSD authoring effort (793 → 300 person-hours), statistically significant on a paired test, at 74.3% reviewer-assessed accuracy and 3.7/5 developer satisfaction. Third, it contributes comparative, practice-based observations on model choice (GPT-4.1/5 versus Claude Sonnet 4) and interaction mode (Ask versus Agent). Fourth, it documents honest limitations, the monotonic decay of benefit with complexity, the sections that resisted automation, the failure modes encountered, and the mitigations that worked, yielding actionable guidance for other organizations.

The remainder of the paper is organized as follows. Section II positions the work in the literature. Section III describes the industrial context, the study interfaces, and the tooling. Section IV presents the framework. Section V reports the evaluation method and results. Section VI discusses interpretation, generalizability, and threats to validity, and Section VII concludes with future work.

## II. RELATED WORK

### A. LLMS IN SOFTWARE ENGINEERING

Transformer-based language models [6] scaled into few-shot learners [7] and trained on code, into practical program synthesizers such as Codex [1] and IntelliCode Compose [8]. Deployed at GitHub Copilot, these models have been evaluated extensively: telemetry-based studies link suggestion acceptance to perceived productivity [2]; a controlled experiment found developers with Copilot completed a standardized task 55.8% faster [3]; and qualitative studies characterize how programmers accept, repair, and steer suggestions [9], [10], [11]. Systematic reviews map a rapidly growing design space of LLM applications across the development lifecycle [4], [12]. Almost all of this literature, however, treats source code, tests, or short in-code artifacts as the unit of generation; multi-page, formally reviewed technical specification and design documents remain at the margins.

### B. AUTOMATED SOFTWARE DOCUMENTATION

Documentation has always been a core part of software engineering, typically written after the code as a summary or guide for developers. LLMs have increasingly been used to generate these documents that were previously produced manually, building on a longer line of work in comment generation, subroutine summarization, and docstring generation [13]–[16]. In every case, the input is code that already exists. This paper does the reverse, generating the document before the code exists from a functional specification, organizational standards, and a mandated template, complete with diagrams, audit tables, and compliance sections. To our knowledge, empirical accounts of LLM-generated, template-governed specification documents adopted in production delivery are rare, and this gap motivates the present study.

### C. PROMPT ENGINEERING AND CHAIN-OF-THOUGHT

Chain-of-thought prompting is known to improve multi-step task performance [5], [7], including in zero-shot settings [17], and has been catalogued for software-engineering tasks [18], [19]. This paper operationalizes it at document scale: rather than a single prompt that "writes the TSD," the task is decomposed into an explicit sequence of grounded reasoning stages, exemplar analysis, template mirroring, context extraction, section-wise generation, diagram generation, and validation, each with controlled inputs and outputs.

### D. INDUSTRIAL STUDIES OF AI-ASSISTED DEVELOPMENT

Industrial evidence on AI pair programming is dominated by code-centric metrics [2], [3], and practitioner-oriented syntheses caution that enterprise adoption requires governance, review discipline, and realistic expectations [20]. Case-study methodology in software engineering emphasizes rich context, explicit metrics, and candid threats-to-validity reporting [21]; we follow that guidance. The study also differs from most published Copilot evaluations in that the generated artifacts were not experimental by-products: the TSDs produced here were reviewed, approved through the organization's standard checklist, and adopted as the delivery documents of record.

## III. STUDY CONTEXT

### A. ORGANIZATIONAL SETTING

The client organization is a large Australian wholesale distributor supporting more than 36,000 customers across 2,500+ retail sites in food, liquor, and hardware. An enterprise-wide transformation program replaces a heavily customized Enterprise Resource Planning (ERP), AS/400 systems, and fragmented TIBCO-based integrations with Microsoft Dynamics 365 and an Azure Integration Services platform (Logic Apps, Azure Functions, Service Bus, Event Grid, Application Programming Interface (API) Management, Azure Data Factory). Delivery is executed with a consulting firm that is a Microsoft solutions partner, whose proprietary Azure-native observability platform, referred to throughout as the System Management Console (SMC), provides end-to-end traceability, observability, business activity monitoring, and self-service failure handling. SMC matters for this study because every TSD must specify the interface audit, logging, and exception-handling contract with SMC; these sections are part of what the framework must generate.

### B. THE COPILOT ADOPTION PROGRAM

Between July 2 and October 24, 2025, the delivery organization ran a structured GitHub Copilot proof-of-concept spanning sixteen use cases in three milestones, covering greenfield and brownfield development of Function Apps, Logic Apps, and Azure Data Factory (ADF) pipelines, code optimization, security and compliance auditing, unit-test generation with Azure DevOps integration, and, the focus of this paper, TSD generation in two maturity levels ("Basic" and "Advanced"). Each use case passed through a governed lifecycle: development, documentation, knowledge-transfer demonstration, and formal approval. Effort, accuracy, and satisfaction were recorded per use case in a shared time-saving workbook, and the prompts for every use case were archived in a dedicated repository (Section IV-E). TSD-generation results were demonstrated to stakeholders in September and October 2025, with a consolidated presentation of the prompt-engineering approach in January 2026.

### C. THE ROLE OF THE TSD

In this organization, the TSD is authored in Confluence against a mandated template of fifteen top-level sections whose contents define the seventeen content areas tracked throughout this study; the full template contents appear in Fig. 1.

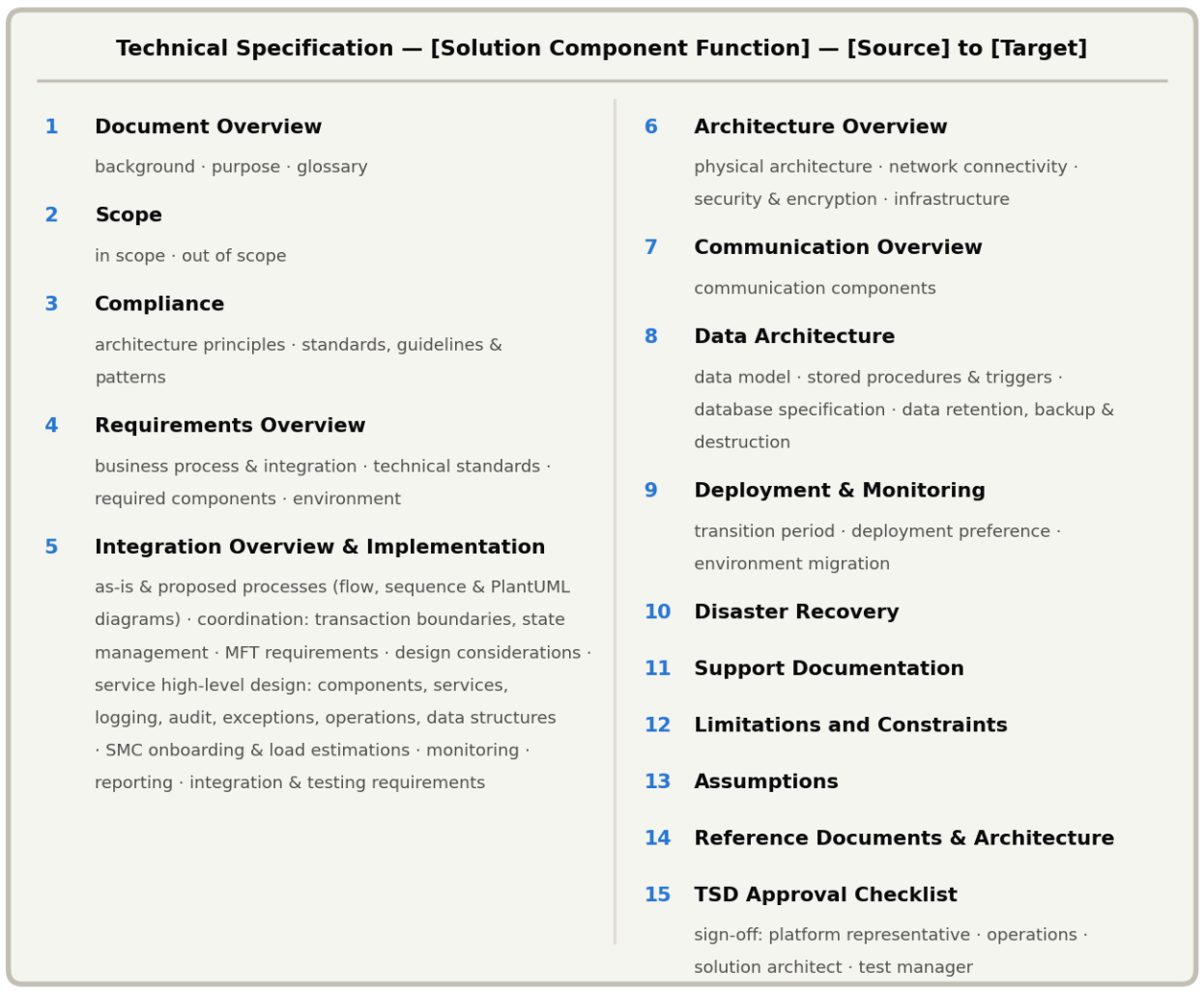

**FIGURE 1. Structure of the organization's mandated Confluence TSD template (anonymized). Its fifteen top-level sections, with the substantial service high-level design and integration-and-testing subsections tracked separately, define the seventeen content areas measured in Section V.**

A few sections convey what the document is required to contain. The integration overview and implementation section specifies the as-is and proposed business processes, carries the process-flow and sequence diagrams, and covers coordination and audit, logging, and exception handling. The service's high-level design and integration and testing requirements are substantial subsections of that section and are tracked separately for measurement. The compliance section binds the design to architecture principles, organizational standards, and required components, and a formal approval checklist for gate sign-off. Across these areas the document must reproduce concrete engineering detail, endpoint tables, retry policies, transaction boundaries, PlantUML sequence definitions, and SMC audit-observation tables, rather than summary prose, which is precisely what makes it expensive to write and a demanding target for generation.

### D. STUDY INTERFACES AND SAMPLING

The framework was developed on interface EAII-270 ("Client Identity"), a greenfield durable Azure Function App managing client-identity application workflows against Microsoft Entra ID app registrations and refined on EAII-265 ("Employee Info Feed"), a greenfield Azure Logic App that delivers daily employee master data from the payroll system to a third-party safety-and-risk platform over Secure File Transfer Protocol (SFTP), Blob Storage, Event Grid, and Service Bus. That deliberate contrast, an API/event-driven compute service versus a file-based data feed, established that one prompt framework could span the organization's two dominant integration styles.

It was then applied to a further thirteen interfaces drawn from the organization's catalogue of over 700, giving a comprehensive set of fifteen documents. Selection was stratified rather than opportunistic: candidates were chosen to span Azure resource type (Function App, Logic App, Azure Data Factory, SQL, and Managed File Transfer (MFT) hybrids), complexity band (Low through Very High on the organization's own rating), criticality tier (Gold, Silver, Bronze), processing style (pass-through, mapping-only, transformation, orchestration), development mode (greenfield and brownfield), and business domain (retail, corporate, B2B trading, supply chain, logistics, marketing analytics, and shared enabling services). Because generalizability is the first objection to any single-organization case study, the sample was constructed so that coverage could be tabulated rather than asserted; Table I summarizes it. Appendix A (Table IV) lists each interface's characteristics individually. Interfaces were selected before their outcomes were known, and no document was excluded after generation.

TABLE I
COMPOSITION OF THE FIFTEEN-DOCUMENT STUDY SAMPLE

| Dimension | Levels represented | Documents (n) |
|---|---|---|
| Azure resource type | Function App | 4 |
| | Logic App | 6 |
| | Azure Data Factory | 3 |
| | SQL | 1 |
| | MFT/file hybrid | 4 [a] |
| Complexity band | Low | 1 |
| | Medium | 4 |
| | High | 6 |
| | Very High | 4 |
| Criticality tier | Gold | 10 |
| | Silver | 4 |
| | Bronze | 1 |
| Processing style | Pass-through | 1 |
| | Mapping-only | 2 |
| | Transformation | 11 |
| | Orchestration | 1 |
| Development mode | Greenfield | 13 |
| | Brownfield | 2 |
| Business domain | Retail | 4 |
| | Corporate | 2 |
| | B2B Trading | 4 |
| | Supply Chain (NZ) | 2 |
| | Logistics | 1 |
| | Marketing Analytics | 2 |
| | Enabling Services | 1 [b] |
| Baseline provenance | Tracked actuals | 11 |
| | Expert Estimates | 4 |

[a] Hybrid interfaces are counted under both contributing resource types, so this dimension totals more than fifteen.
[b] One interface spans two business domains and is counted under both.

Complexity band is the organization's own rating. Criticality tier is the proposed tier from the interface catalogue. Brownfield cases had code already in place. Baseline provenance is analyzed separately in Section V-C.

### E. TOOLING

All generations were performed with GitHub Copilot inside the developers' standard Integrated Development Environments (IDEs) (Visual Studio and Visual Studio Code), using repository-level guidance files (copilot-instructions.md) to carry standing conventions [22]. Copilot's model picker exposed OpenAI's GPT-4.1 and GPT-5 [23], and Anthropic's Claude Sonnet 4 [24], and both of Copilot's interaction styles were exercised: Ask mode (turn-by-turn conversational prompting) and Agent mode (autonomous multi-step execution against the workspace). No fine-tuning, retrieval infrastructure, or bespoke orchestration software was used; the framework is purely a prompt-engineering and process discipline over commercially available tooling embedded in the delivery environment (Fig. 2), which is central to its practical transferability.

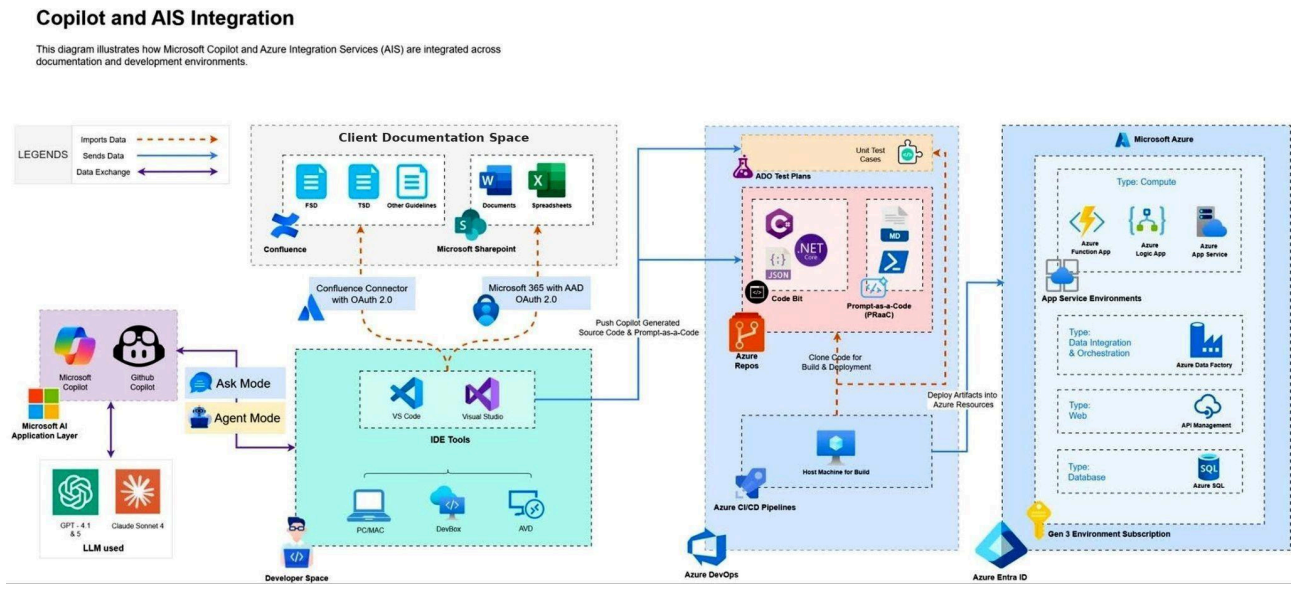

**FIGURE 2. Integration of GitHub Copilot with the AIS delivery environment, anonymized from the project's architecture diagram. Documentation sources flow through the IDE-based Copilot workspace (Ask and Agent modes over GPT-4.1/5 and Claude Sonnet 4) into the version-controlled Prompt-as-a-Code repository, Continuous Integration / Continuous Delivery (CI/CD) pipelines, and the Azure platform.**

## IV. A CHAIN-OF-THOUGHT FRAMEWORK FOR TSD GENERATION

### A. *FROM SINGLE-PASS PROMPTING TO A PHASED DESIGN*

The phased design was not the starting point. The team's earliest attempts used single-pass prompting: one comprehensive prompt asking Copilot to produce a complete TSD directly from the FSD. These attempts exhibited the failure modes anticipated in Section I: plausible but generic content, silently skipped template sections, and unstated design assumptions. Published evidence pointed toward decomposition as the remedy. Chain-of-thought prompting is the practice of asking a model to work through a problem in explicit intermediate steps rather than jumping straight to a final answer. Eliciting those steps substantially improves performance on tasks that require multiple stages of reasoning [5], [17]. Prompt chaining applies the same principle across calls rather than within one: a complex task is decomposed into a sequence of smaller prompts, each taking the previous output as input. Studies report better output quality, controllability, and error isolation than a single monolithic prompt [25], and the approach is now codified in practitioner guidance [26]. On this basis, the team moved to a multi-phase prompting strategy that separates structure generation, content population, and validation, and all work reported in the remainder of this paper uses that design.

### B. *OVERVIEW AND DESIGN PRINCIPLES*

The framework rests on four principles distilled during its development.

1) *Context building:* before any content is requested, the model is grounded in a fixed knowledge base of organizational material, the mandated template, the architecture standards and naming conventions, and approved FSD/TSD exemplar pairs from prior interfaces. Grounding here means supplying that material to the model as explicit context in the opening stages, so that subsequent generation is conditioned on the organization's own conventions rather than on generic notions of what a design document contains. The exemplars establish expected structure, tone, and level of detail, and demonstrate how business requirements translate into technical specifications.
2) *Structured phase generation:* the work is decomposed into phases with explicit, narrow goals, document structure, scope interpretation, and data-flow extraction, so the model concentrates on one task at a time instead of attempting everything at once.
3) *Incremental assembly:* later stages reuse and extend validated earlier outputs, so the model refines the document rather than rewriting it or hallucinating missing details.
4) *Iterative refinement through feedback loops:* intermediate human review checkpoints validate each generated increment against business logic, clarity, and technical consistency before the next stage proceeds.

Fig. 3 shows the resulting architecture; Fig. 4 shows the stage pipeline.

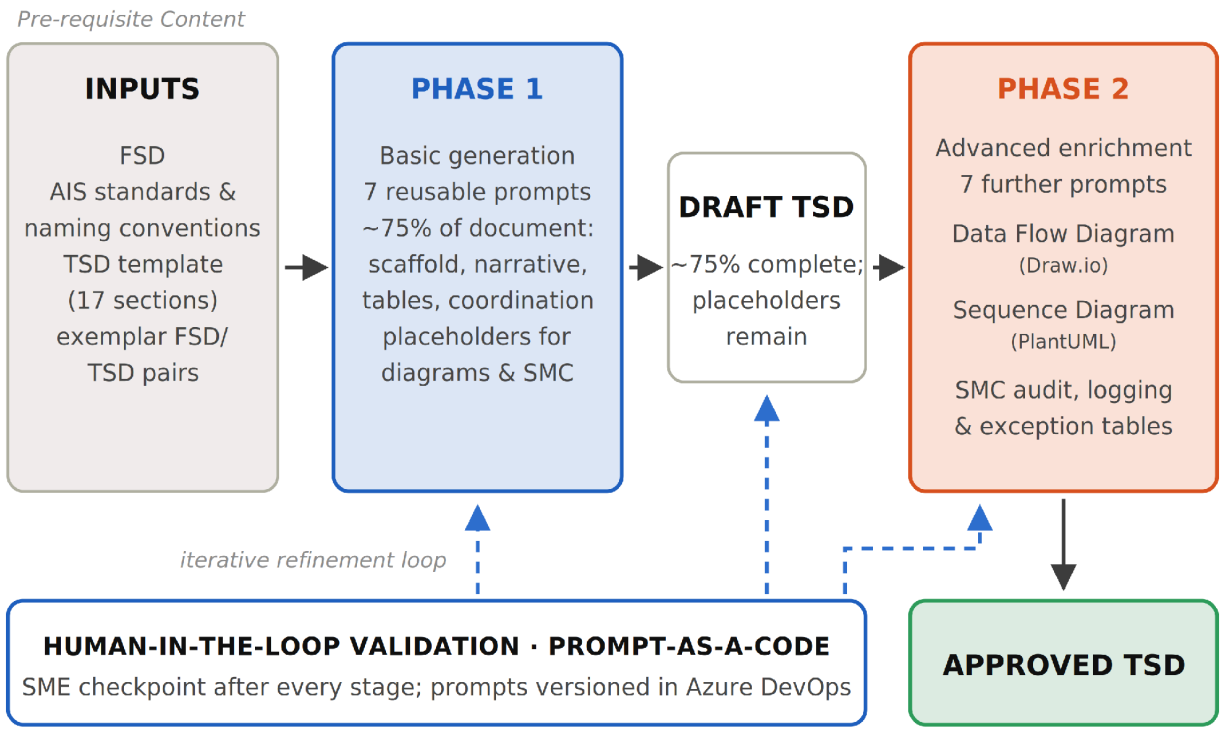

**FIGURE 3. Two-phase TSD generation architecture. Phase 1 grounds the model in the FSD, organizational standards, and the approved template to generate ~75% of the document; Phase 2 enriches the draft with diagrams and SMC audit/logging specifications; human validation**

**checkpoints gate both phases, and every prompt is versioned in the Prompt-as-a-Code repository.**

### C. PHASE 1, BASIC GENERATION (~75% OF THE DOCUMENT)

Phase 1 consumes four grounding inputs: the interface's FSD; the organization's AIS standards and naming conventions; the approved Confluence TSD template; and exemplar FSD/TSD pairs from previously delivered interfaces. Guided by reusable prompts, Copilot mirrors the mandated template (Fig. 1). The Phase 1 draft populates approximately three quarters of the template's content areas, specifically: background and purpose, scope and constraints, compliance references, requirements overview, the proposed process description, coordination semantics (transaction boundaries, retries, idempotency, state management), security and infrastructure sections, data-architecture and transformation rules extracted from the FSD mapping, testing strategy, deployment and monitoring considerations, disaster recovery, limitations, and assumptions. The remaining areas, those requiring visual artifacts or observability contracts, are output as explicit placeholders rather than invented, a deliberate control against hallucination.

### D. PHASE 2, ADVANCED ENRICHMENT (~25% OF THE DOCUMENT)

Phase 2 completes the draft with the content that the initial single-pass attempts handled sub-optimally. Further prompts direct Copilot to:

1. generate process-flow and physical-architecture diagrams through Draw.io automation;
2. produce PlantUML sequence definitions capturing the end-to-end message flow, including retry and compensation paths; and
3. populate the SMC audit, logging, and exception-handling tables, observation stages, reference keys, error types and codes, and recovery actions, in the tabular form the template mandates.

Because these artifacts depend on design decisions fixed in Phase 1, sequencing them second lets the model derive them from an already-validated textual design instead of speculating one.

### E. THE SIX-STAGE COT PIPELINE

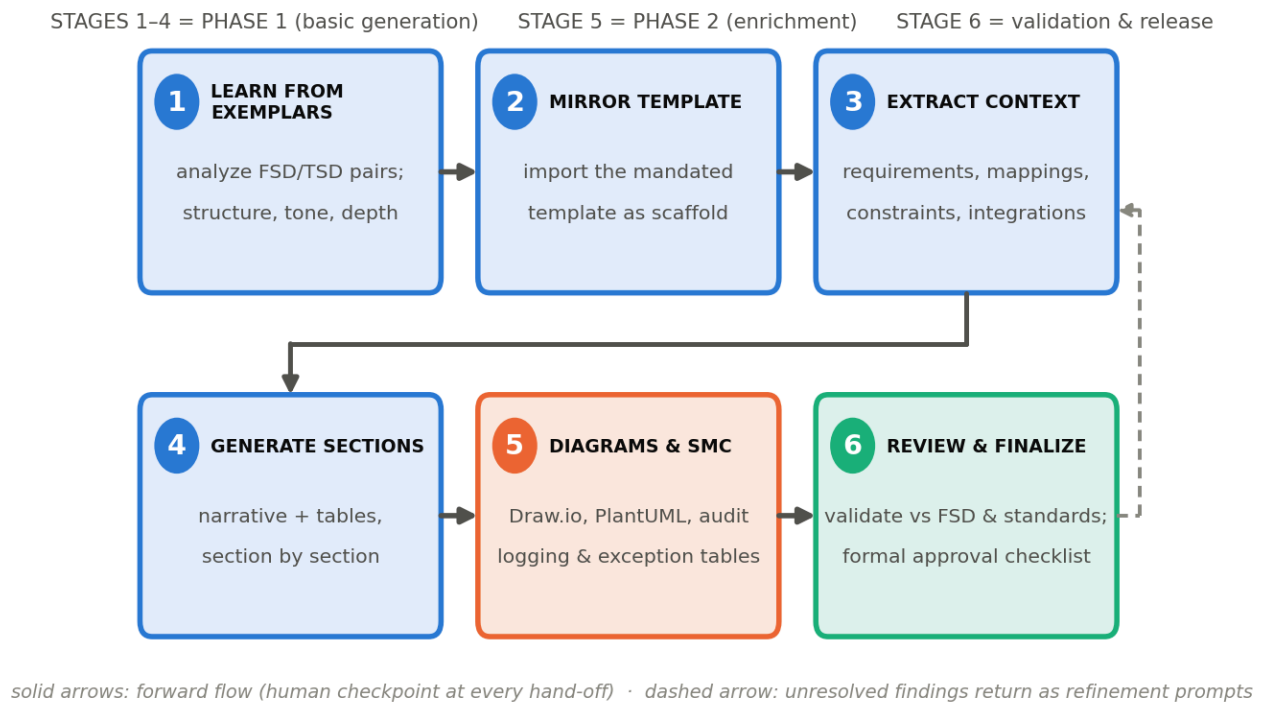

**FIGURE 4. The six-stage chain-of-thought pipeline. Stages 1–4 constitute Phase 1, stage 5 constitutes Phase 2, and stage 6 closes the loop; a human validation checkpoint follows every stage, and unresolved review findings return to earlier stages as refinement prompts.**

Operationally, the phases unfold as six chained stages (Fig. 4):

1. *Learn from exemplars:* analyze approved FSD/TSD pairs to internalize structure, tone, and depth.
2. *Mirror the template:* import the mandated template as the scaffold for all subsequent generation.
3. *Extract context:* pull requirements, mappings, constraints, and integration points from the FSD.
4. *Generate sections:* produce narrative and tabular content section by section.
5. *Create diagrams and SMC tables:* generate Draw.io and PlantUML artifacts and the audit, logging, and exception tables.
6. *Review and finalize:* validate against the FSD and organizational standards and complete the formal approval checklist.

Stages 1–4 constitute Phase 1; stage 5 constitutes Phase 2, and stage 6 closes the loop. Each stage's prompt states its goal, its permitted inputs (including prior stage outputs), and the exact output format expected. This explicit decomposition is what makes the approach chain-of-thought at document scale: instead of one opaque leap from FSD to finished TSD, the model is walked through requirement understanding, template mapping, content generation, and validation as separate reasoning steps [5]. In the team's experience, this directly attacked two recurring failure modes of single-pass prompting, silently skipped instructions and unstated assumptions, because each narrow stage leaves the model less scope to omit or invent, and each checkpoint catches drift while it is still cheap to correct. A representative four-prompt sequence from the team's proof-of-concept demonstration is reproduced in Appendix B.

### F. PROMPT-AS-A-CODE (PraaC)

Every prompt in the pipeline is treated as an engineering artifact. Prompts are stored in a dedicated Azure DevOps Git repository, organized by artifact type (e.g., a TSD folder containing the per-interface prompt files), versioned, code-reviewed, and evolved through pull requests exactly like source code. Because the repository is hosted in Azure DevOps, GitHub Copilot can reference the prompt files directly through its native integration, so an engineer beginning a new interface retrieves the current version of a prompt in the development environment rather than copying it from an external document. Companion "prompts-and-outcomes" documents record each prompt alongside representative outputs for auditability. PraaC yielded two practical benefits. First, transferability: the prompt set engineered for EAII-270 was adapted to EAII-265, a different Azure resource type and integration pattern, with refinement rather than redesign, demonstrating that a common prompt structure can be reused across interfaces while preserving accuracy and consistency. Second, institutional memory: improvements discovered on one interface (for example, wording that prevents the model from truncating long mapping tables) propagate to every subsequent interface through the shared repository.

### G. HUMAN-IN-THE-LOOP VALIDATION

The framework assigns labor explicitly. Copilot drafts boilerplate and structured sections, aligns content with the Confluence template, and iterates quickly; human experts design the overall TSD structure, resolve complex logic, own diagrams' semantic correctness, and uphold documentation accuracy. Validation checkpoints follow every stage: generated sections are compared against the FSD and against previously approved, manually authored TSDs; technical material is verified with the team lead; and flowchart prompts are iterated until diagrams meet professional standards for iconography and vocabulary. The finished artifacts are explicitly labeled as Copilot-plus-manual products and pass the organization's standard TSD approval checklist before release; the framework automates drafting, not accountability.

## V. EVALUATION

### A. METHOD AND METRICS

For every document the team recorded six measures:

1. *Framework effort*: person-hours to produce the TSD with the framework, split by phase.
2. *Manual baseline*: person-hours to author an equivalent document by hand.
3. *Accuracy*: the share of generated content accepted without correction during validation.
4. *Template compliance*: the percentage of mandated sections and subsections present in the generated document. Compliance measures presence; the adequacy of what each section contains is captured by accuracy.
5. *Developer satisfaction*: rated on a five-point scale.
6. *Section rework:* section by section, whether each of the seventeen tracked template sections required manual rework.

Baselines were paired with documents at the interface level: the manual baseline for a given interface is the effort recorded or estimated for the manually authored TSD of that same interface, not a category or complexity-band average. Eleven of the fifteen baselines are tracked actuals taken from the delivery team's time records for that interface. The remaining four (EAII-215, EAII-249, EAII-260, and EAII-436) are expert estimates, supplied by the delivery developer who authored the corresponding manual TSD, in cases where historical time records were no longer retrievable; these are marked with asterisks in Fig. 5. Estimates were given as a single figure for total authoring effort rather than reconstructed per phase. Section V-C reports the analysis restricted to the eleven tracked baselines.

Reviewer assessment was performed by the delivery developers responsible for the corresponding manually authored TSDs, so generated output was judged directly against the organization's own established practice by the people who know it best. Baseline provenance was recorded explicitly: eleven of the fifteen baselines are tracked actuals from time records in Azure DevOps (ADO) Boards, and four are expert estimates made where historical actuals were unavailable in ADO. Because estimated baselines are the softer evidence, every headline result below is also reported for the recorded-only subset. All fifteen generated TSDs were reviewed, corrected where needed, passed through the organization's standard TSD approval checklist, and adopted as delivery artifacts of record.

Effort reductions are paired observations; each document has a manual and a framework value, so we test them with the Wilcoxon signed-rank test, which makes no normality assumption and is appropriate at this sample size. Relationships between interface size and outcome are reported as Spearman rank correlations. We report effect sizes and distributions rather than significance alone, since with fifteen documents the practical question is how large and how variable the benefit is, not merely whether it is non-zero [27].

### B. EFFORT REDUCTION (RQ1, RQ3)

Across the fifteen documents, TSD authoring effort fell from 793 to 300 person-hours: 493 hours saved, a pooled reduction of 62.2%. The mean per-document reduction was 64.0% (median 64.0%, standard deviation 6.0 percentage points) and every document in the sample improved, with reductions ranging from 50.0% on the most complex interface to 72.5% on a mid-sized Azure Data Factory use-case.

The 300 hours cover per-document production effort using the established prompt set. One-time investments in prompt engineering, exemplar curation, and repository setup preceded the measured period and are excluded. The reductions reported here therefore describe steady-state effort under reuse, not the cost of adopting the framework from scratch.

The paired difference is statistically significant (Wilcoxon signed-rank, $n = 15$, $W = 0$, $p = 0.00065$). The same test restricted to the eleven documents with tracked rather than estimated baselines gives a mean reduction of 64.5% ($p = 0.00098$), so the finding does not depend on the four estimated cases. Fig. 5 shows the per-document picture and Table II reports the distribution.

Within the 300 framework hours, enrichment (Phase 2) accounted for roughly 61%, consistent with diagrams and observability content remaining the costliest part of the document to produce.

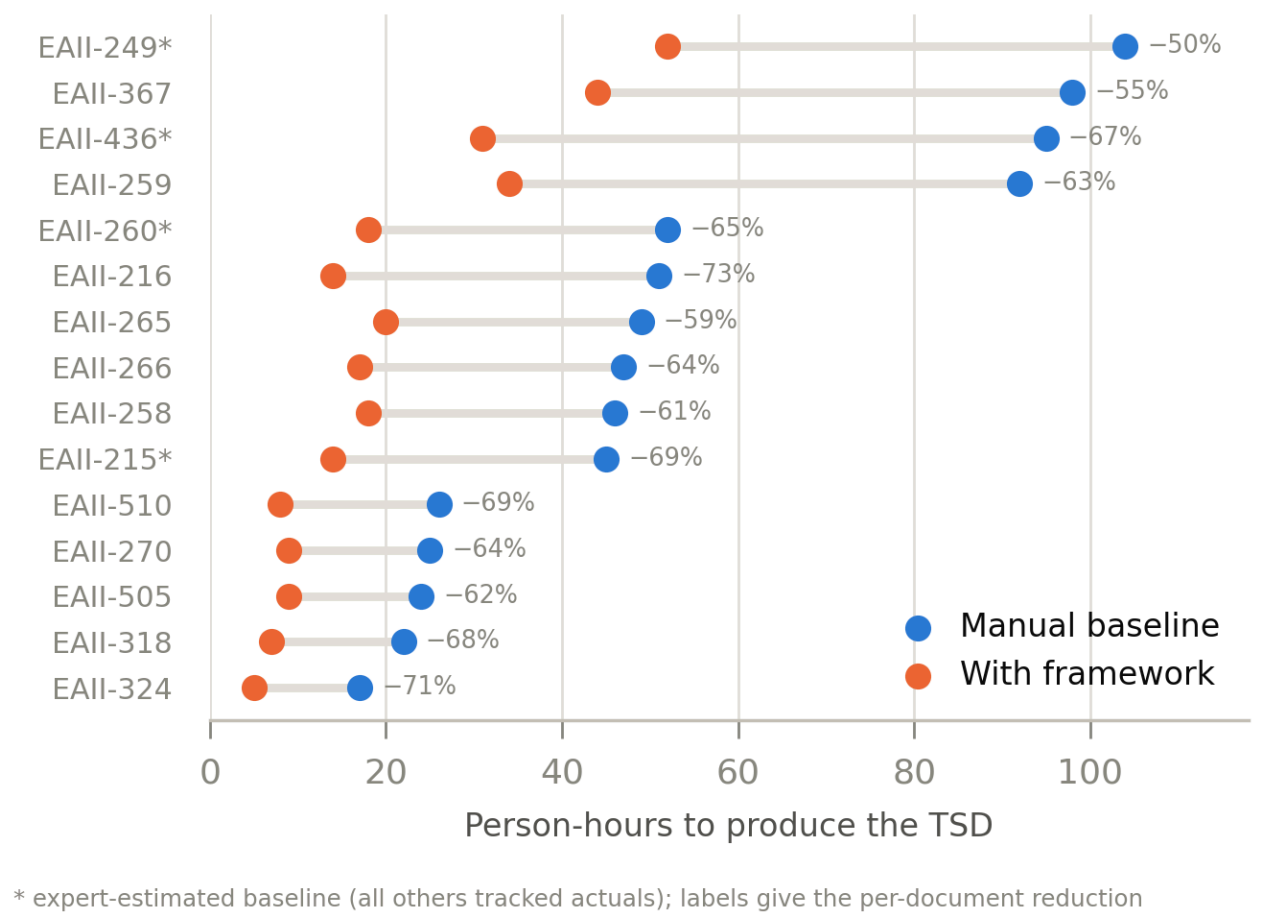

**FIGURE 5. Manual baseline versus effort with the framework for each of the fifteen documents, ordered by baseline size. Every document improved; labels give the per-document reduction, and asterisks mark the four expert-estimated baselines.**

TABLE II
EFFORT, ACCURACY, AND SATISFACTION ACROSS THE FIFTEEN DOCUMENTS

| Measure | Mean | Median | Range | All 15 / Recorded 11 |
|---|---|---|---|---|
| Manual baseline (h) | 52.9 | 47.0 | 17–104 | 793 h total |
| Effort with framework (h) | 20.0 | 17.0 | 5–52 | 300 h total |
| Effort reduction (%) | 64.0 | 64.0 | 50.0–72.5 | 64.0 / 64.5 |
| Reviewer-assessed accuracy (%) | 74.3 | 75 | 55–91 | - |
| Template compliance (%) | 89.0 | 89 | 75–98 | - |
| Developer satisfaction (of 5) | 3.70 | 4.00 | 2.5–4.5 | - |
| Sections requiring rework (of 17) | 2.8 | 2 | 1–6 | - |

Pooled reduction across all documents is 62.2% (793 → 300 h); the mean of the per-document reductions is 64.0%. The two differ because larger documents, which weight the pooled figure more heavily, show smaller proportional gains. Reviewer accuracy is the share of generated content elements accepted without correction during validation. Sections are decomposed into elements, paragraphs, table rows, diagram nodes, and list items, and an element counts as accurate if it survived validation unchanged.

The gap between the pooled 62.2% and the per-document mean of 64.0% is itself informative: the pooled figure is dominated by the largest documents, and those are precisely the ones where the framework helps least. Answering RQ1, phased CoT prompting with validation checkpoints produced standards-compliant TSDs that passed formal approval at roughly a third of the manual cost, consistently and significantly across a deliberately varied sample.

### *C. BENEFIT DECAYS WITH COMPLEXITY (RQ3)*

The most consequential result is not the average but its structure. Sorting documents by the organization's own complexity rating yields a monotonic decline in effort for every outcome measure (Fig. 6): effort reduction falls from 70.6% (Low) through 66.0% (Medium) and 65.1% (High) to 58.9% (Very High); reviewer-assessed accuracy falls from 91% to 81.5%, 74.5%, and 62.5%; and developer satisfaction falls from 4.5 to 3.0.

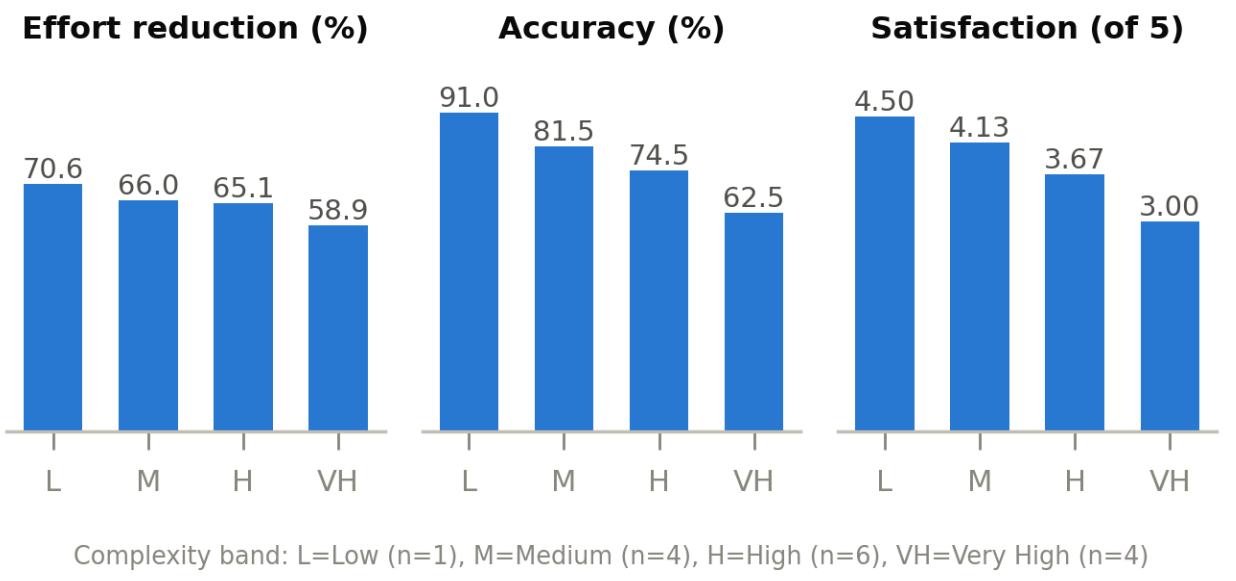

**FIGURE 6. Effort reduction, reviewer-assessed accuracy, and developer satisfaction by interface complexity band. All three decline monotonically as complexity rises.**

Fig. 7 locates each document in the savings–quality plane: the goal quadrant of above-median savings at above-median accuracy is populated only by Low-to-High complexity documents, and no Very High complexity document reaches the desired accuracy median. The same pattern appears against a continuous proxy for scale: larger manual baselines are associated with lower proportional savings (Spearman $\rho = -0.48$, $p = 0.069$) and markedly lower accuracy ($\rho = -0.99$, $p < 0.001$; Fig. 8). The accuracy correlation is strikingly tight and should be read with care, complexity rating and expected document size are not independent judgments in this organization, and both were assessed by the same delivery team, but the direction is unambiguous and matches the qualitative record.

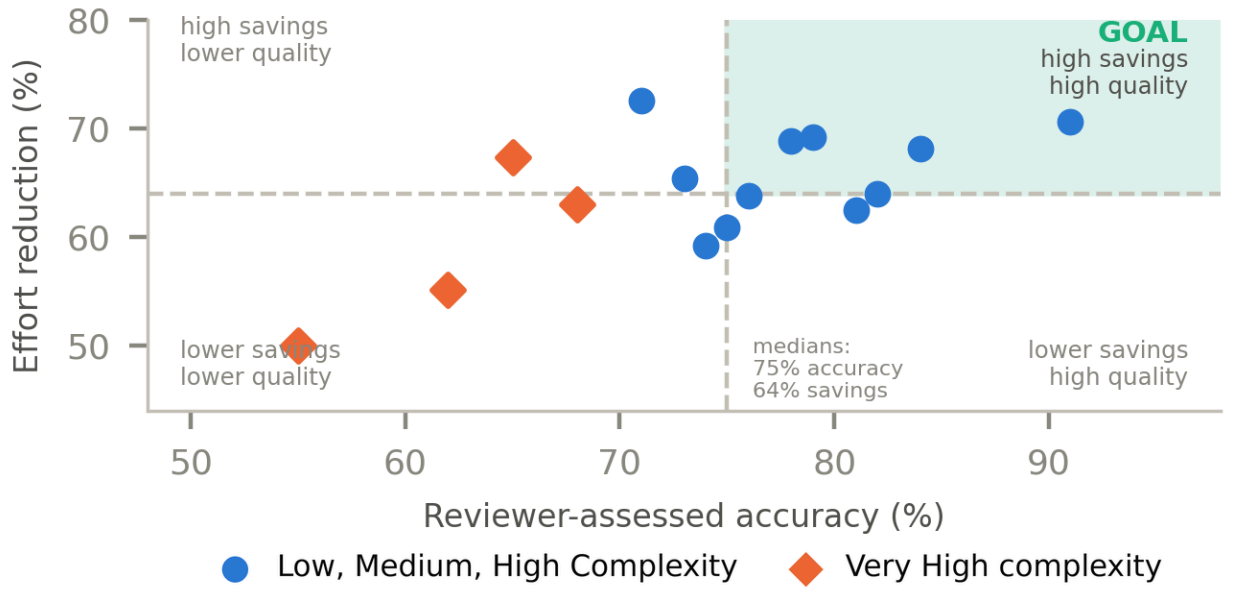

**FIGURE 7. The fifteen documents in the savings–quality plane, partitioned at the sample medians (75% accuracy, 64.0% effort reduction). The shaded quadrant, above-median savings at above-median accuracy, is the operating goal; every Very High complexity document (diamonds) falls short of the accuracy median.**

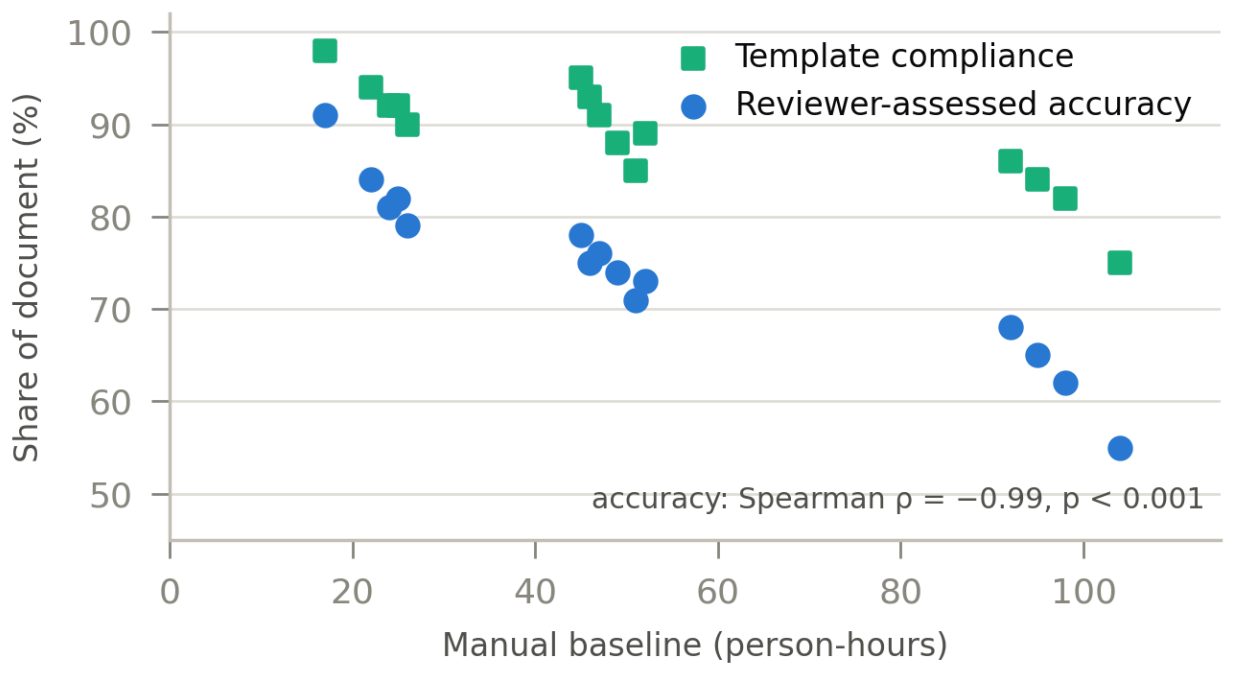

**FIGURE 8. Reviewer-assessed accuracy and template compliance against the manual baseline for each document. Each of the fifteen documents contributes exactly two points: one blue circle for its accuracy and one green square for its template compliance, both plotted at that document's manual baseline on the horizontal axis. Points cluster where baselines are similar, which can make the thirty markers difficult to count. Both measures decline as documents grow; the accuracy correlation (Spearman ρ = −0.99) shares an assessor with the complexity ratings and is read as a strong directional signal rather than a precise estimate.**

This matters for practice more than the headline average does. A reader deciding where to deploy such a framework needs to know that its return is concentrated in the routine half of the portfolio: simple pass-through and mapping interfaces are drafted almost entirely by the model, while Very High complexity integrations still consume forty to fifty person-hours of expert work and yield documents whose generated content is only about three-fifths acceptable as produced. The framework redistributes expert effort toward hard problems; it does not eliminate it.

### D. WHERE THE HUMAN EFFORT GOES

Section-level tracking localizes the residual work. Across the fifteen documents, a mean of 2.8 of the seventeen template sections required manual rework (median 2, range 1–6), and the burden is highly concentrated: sequence diagrams, process-flow diagrams, and architecture overview sections were reworked far more often than any narrative section, followed by service high-level design and the integration and testing material. By contrast, the document overview, scope, requirements overview, communication overview, and data architecture sections were accepted essentially as generated. Reviewers consistently judged the integration and dependency content for the weakest and least certain, because it depends on external systems, evolving interfaces, and cross-team alignment that no single FSD contains. Happy-path flows were well covered; boundary conditions, failure scenarios, and recovery flows required manual expansion. The rework burden also scales with complexity, rising from a mean of 1.0 section on the Low band to 4.75 on Very High (Fig. 9).

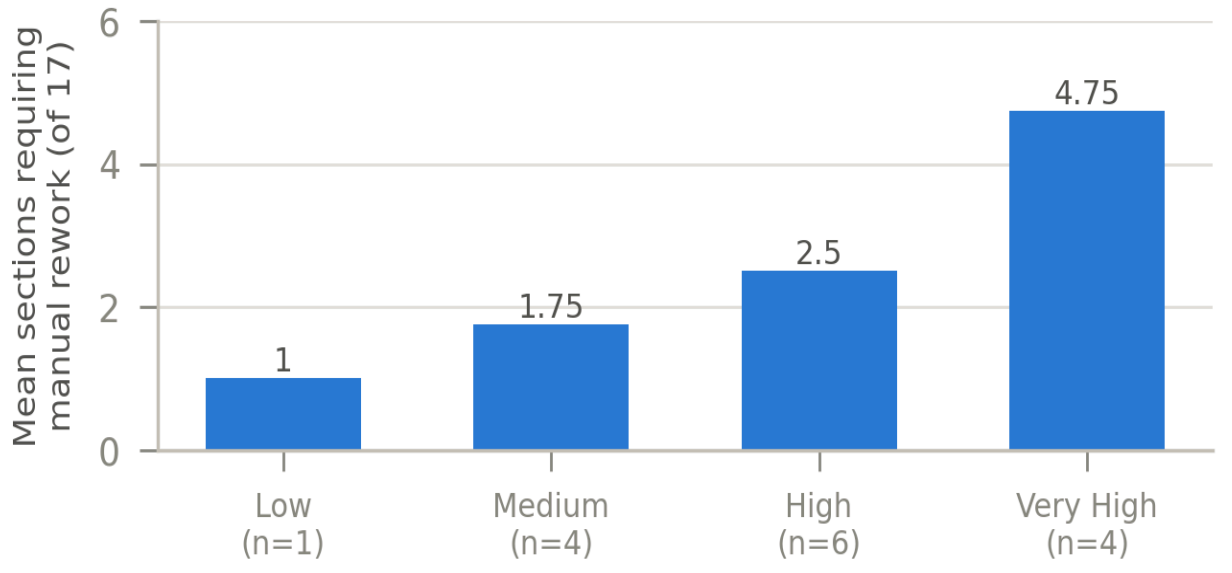

**FIGURE 9. Mean number of template sections (of seventeen) requiring manual rework, by interface complexity band.**

A complementary view comes from the delivery-effort breakdown recorded for the two originating interfaces (Fig. 10). The figure divides each interface's total framework effort into activities. Prompt engineering was the single largest activity, taking 62% of total effort on the Function App interface and 56% on the Logic App interface, while reviewing and understanding the source specifications took only 7% and 4% of total effort respectively. This is the clearest statement of what the framework does to a delivery team's work: it does not remove expert labor so much as relocate it, from writing and transcribing toward specifying, steering, and verifying. It also explains why Prompt-as-a-Code matters economically. If most of the remaining effort is spent constructing prompts, then prompts are the reusable asset, and versioning them is not administrative tidiness, but the mechanism by which that majority cost is amortized across subsequent interfaces.

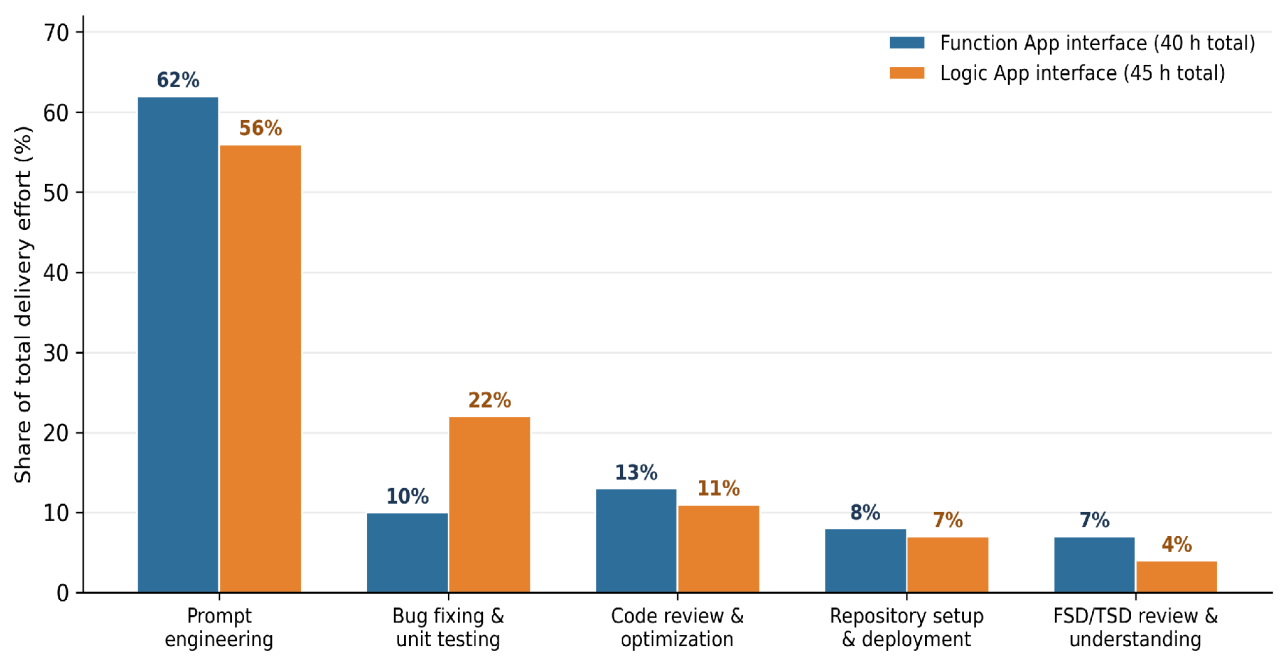

**FIGURE 10. Distribution of delivery effort under the framework for the two originating interfaces. Prompt engineering dominates; specification review shrinks to a small share.**

Three recurring failure modes and their mitigations are worth recording. Early outputs contained generic, best-practice-flavored assumptions; the mitigation was aggressive grounding, refining integration points, workflows, and constraints against the actual target systems and architecture standards. The model intermittently skipped portions of long prompts; CoT decomposition with deliberately concise per-stage prompts largely eliminated this. Raw output needed restructuring for enterprise readiness and Confluence publication, addressed by the template-mirroring stage and a final manual pass. Consistent with the program's observations on generated code, the generated TSDs were praised for structural consistency and uniform technical language across sections.

### E. MODEL AND MODE OBSERVATIONS (RQ2)

Model and mode selection was pragmatic rather than experimental. Developers switched among GPT-4.1, GPT-5, and Claude Sonnet 4 and used both of Copilot's interaction styles as the work demanded; across the sample, eight documents were produced primarily with GPT-4.1/5 and seven primarily with Claude Sonnet 4, and every document used a combination of Ask and Agent mode rather than one exclusively. That usage pattern is a finding in itself; practitioners treated the modes as complementary, using Ask for targeted section work and Agent for multi-step assembly, but it also means the present data cannot isolate a model or mode effect, since neither factor was varied under controlled conditions. Mean effort reduction differed by less than two percentage points between the two model groups, which we report only to note that no meaningful difference is detectable at this sample size and design.

The team's qualitative differentiating observations, recorded during the program, are reproduced in Table III. Each observation was recorded in the context of one Copilot interaction mode and attributed to the model that performed better in that mode. They are experience reports from production use, not controlled measurements, and are presented as hypotheses for the benchmarking work described in Section VII.

TABLE III

TEAM OBSERVATIONS FROM USING TWO MODEL FAMILIES IN TWO INTERACTION MODES

| # | Observation (verbatim from study records) | Copilot mode | Model favored |
|---|---|---|---|
| 1 | Better at understanding long contextual prompts and produces better accuracy | Ask | Claude Sonnet 4 |
| 2 | Better at interpreting ambiguous user queries | Ask | Claude Sonnet 4 |
| 3 | Stronger in natural, human-like explanations and summarization | Ask | Claude Sonnet 4 |
| 4 | Better at understanding the Azure ecosystem | Ask | GPT-4.1/5 |
| 5 | More accurate and context-aware responses for short prompts | Ask | GPT-4.1/5 |
| 6 | Better with third-party developer-tool integration (e.g., SMC) | Agent | GPT-4.1/5 |
| 7 | Better for structured workflows such as unit-test case generation | Agent | GPT-4.1/5 |
| 8 | Better at generating responses autonomously with minimal supervision | Agent | Claude Sonnet 4 |
| 9 | Better for TSD generation | Agent | Claude Sonnet 4 |
| 10 | Better in adaptability and flexibility in multi-turn conversations | Agent | Claude Sonnet 4 |

Attributions are as recorded on the program's model-comparison slide: each observation arose while using the stated Copilot interaction mode and names the model the team found stronger in that context.

### F. PROGRAM CONTEXT

The TSD work sits inside a sixteen-use-case Copilot adoption program covering greenfield and brownfield development, code optimization, security and compliance auditing, and unit-test generation, which recorded effort reductions in the range of roughly 55–75% across those activities with average accuracy near 82% and average satisfaction near 4.3/5. We report this only as context for the delivery environment. The program's own TSD figures have been superseded by the direct measurements presented here and are not carried forward, and the non-TSD activities were measured under a different protocol; no comparison between them and the present results is intended.

## VI. DISCUSSION

### A. WHY PHASED GENERATION WORKS

The COT prompting strategy, which separates structure generation, content population, and validation, produces more accurate and consistent TSDs than single-pass prompting. Three mechanisms motivate it.

1. *Decomposition narrows scope.* A single prompt asking for a complete TSD forces the model to trade the template's fifteen sections off against one another, and the weaker sections are the ones silently dropped. Each stage instead asks for one bounded product against a fixed part of the template, so no section can be quietly sacrificed to finish another. This is the same effect chain-of-thought produces on multi-step problems [5] and that chained prompts produce through error isolation [25].

2. *Grounding precedes generation.* The organization's standards, template, and exemplar TSD pairs are largely tacit knowledge held by the delivery team. The early stages convert that knowledge into explicit context before any prose is requested, so the model writes against the organization's conventions rather than a generic notion of what a design document looks like.
3. *Incremental assembly compounds validation.* Stage n consumes the human-approved output of stage n−1, so an error is corrected while it is still confined to one section rather than propagating into every artifact derived from it. This is why the diagrams and SMC audit tables are generated last: they are derived from a design a human has already accepted, not co-invented alongside it.

The study's own evidence is consistent with this account. The team's early single-pass attempts exhibited precisely the failure modes the prompting literature predicts for monolithic prompts (Section IV-A), and they disappeared under decomposition. The section-level results point the same way: the sections that survive generation untouched are precisely those the template and FSD determine most tightly, while the reworked ones are those requiring design decisions the knowledge base does not contain.

### B. RELATION TO PRIOR FINDINGS

A 62% pooled effort reduction on documentation is of the same order as, and somewhat larger than, the 55.8% speed-up reported for a controlled coding task with Copilot [3]. We do not interpret this as models being better at documents than code; the accuracy figures point the other way, but as template-governed documentation being unusually amenable to grounded generation: much of a TSD's cost is faithful transcription of decisions into a mandated structure, exactly the template-conformant generation that models do well when the structure is supplied. The corollary, visible in the 74% mean accuracy and in the complexity gradient, is that the residual third of the effort is not noise but the irreducible design content of the document, and that its share grows precisely as the engineering problem gets harder.

### C. PRACTICAL GUIDANCE

For organizations attempting the same, four transferable practices emerge. Treat prompts as code: version them, review them, and reuse them across interfaces; PraaC is what turned a one-off experiment into a repeatable capability. Invest in the knowledge base before prompting: the template, standards, naming conventions, and exemplar pairs did more for output quality than refinements to prompt phrasing. Expect the return to concentrate in the routine part of the portfolio, and plan capacity accordingly rather than applying a single average saving across all interfaces. Generate placeholders, not guesses instructing the model to output explicit placeholders for diagrams and unverifiable facts converts hallucination risk into a visible worklist. Plan for a persistent human share that scales with difficulty: budget roughly a third of prior effort on routine interfaces and closer to half on the most complex ones, concentrated in diagrams, sequence flows, and cross-system dependency analysis, and route it through defined checkpoints rather than a single end review.

### D. GENERALIZATION

Nothing in the framework is specific to Azure Integration Services: its inputs are a requirements document, organizational standards, a mandated template, and examples, all of which exist for API specifications, operations runbooks, test plans, and architecture decision records in most delivery organizations. Transfer across fifteen interfaces spanning four Azure resource types, four complexity bands, and seven business domains, with every document improving, is reasonably strong evidence of within-organization generality. Cross-organization generality is untested: template depth, review culture, and the availability of high-quality examples vary, and the framework's dependence on a strong knowledge base means organizations with weak documentation baselines may see smaller gains.

### E. THREATS TO VALIDITY

*Construct validity:* Accuracy is an element-level count: each paragraph, table row, diagram node, and list item either survived validation unchanged or did not. The count itself is objective, but the judgment of what required correction was made by the delivery developers who authored the corresponding manual TSDs. Those reviewers were not blinded and had a stake in the outcome, so the threat lies in the correction decisions rather than in the tallying. The complexity ratings were assigned by the same developers, so accuracy and complexity share a source. A developer who judged an interface complex may also have applied a stricter correction standard to its generated content, producing a correlation between the two even without any real relationship. The observed correlation ($\rho = -0.99$) is therefore almost certainly inflated by that shared source and should be read as a direction rather than a magnitude: accuracy declines as complexity rises, but the strength of that relationship is not established by this design. Four of the fifteen manual baselines are expert estimates rather than tracked actuals, though the principal result holds the recorded-only subset. The team both built and evaluated the framework; reviewers were not blinded; learning effects across the program may inflate later documents' efficiency, and the sample was not randomized in time. Satisfaction is self-reported. *External validity*: The evaluation covers fifteen interfaces in one organization on one toolchain (GitHub Copilot with GPT-4.1/5 and Claude Sonnet 4, 2025–2026); model capabilities are moving quickly, and results may not transfer to organizations without a mandated template and a knowledge base of approved exemplars. *Conclusion validity*: The Wilcoxon test establishes that the paired reduction is not attributable to chance, but with fifteen documents in one estate, the stratum-level figures, particularly the single Low complexity and single Bronze criticality cases, are indicative only, and the complexity gradient should be treated as a strong signal awaiting replication. The adoption of phased over single-pass prompting rests on the team's early production experience and on prior published evidence [5], [25], [26]; quantifying that difference under controlled conditions in this setting remains open. The model and mode observations are practice reports, not benchmarks.

## VII. CONCLUSION AND FUTURE WORK

This industrial case study showed that a chain-of-thought prompt-engineering framework, two-phase generation, a six-stage pipeline with human validation checkpoints, and version-controlled Prompt-as-a-Code enabled a commercial AI coding assistant to produce standards-compliant Technical Specification Documents that were formally approved and adopted in a large enterprise integration program. Across a stratified sample of fifteen production interfaces, authoring effort fell from 793 to 300 person-hours (62.2% pooled; 64.0% mean per document; $p < 0.001$) at 74.3% reviewer-assessed accuracy and 3.7/5 developer satisfaction, with every document in the sample improving. The benefit is real but not uniform: it decays monotonically with interface complexity, from roughly 71% effort reduction and 91% accuracy on the simplest interface to 59% and 62.5% on the most complex, and the residual human effort concentrates in diagrams, sequence flows, and cross-system dependency content. The framework's value lies less in any single prompt than in its process discipline: grounding before generation, decomposition over single-pass ambition, placeholders over guesses, and checkpoints over end-stage review.

The project's recorded roadmap defines future work. Nearest-term publishing automation: integrating generation with Confluence and SharePoint to auto-publish TSDs, synchronize versions, and maintain a single source of truth. Generation to Confluence-ready form is demonstrated, but publication automation was outside the evaluated scope. Second is deepening coverage of the weakest sections: prompting the model to proactively generate dependency maps and risk summaries earlier in design, standardized assumption-validation checklists, and systematic edge-case elicitation (boundary conditions, failure scenarios, recovery flows). Third is lifecycle integration, connecting generation to Azure DevOps and Jira so specifications pull user stories and requirement changes automatically, and ultimately self-updating documentation that reflects code changes in real time. Finally, two specific experiments would strengthen the evidence base. First, independent multi-rater scoring by reviewers who did not author the prompts, reported with inter-rater agreement, which would decouple the accuracy measure from the team that produced the documents and allow the complexity gradient to be re-estimated without a common-assessor confound. Second, a controlled model and mode comparison under identical prompts, converting Table III's practice observations into benchmarked findings. Extending the sample across document types and, ideally, across organizations would then address the external-validity limits that no single-estate study can resolve.

## APPENDIX A: CHARACTERISTICS OF THE FIFTEEN STUDY INTERFACES

Table IV lists the fifteen study interfaces individually across the six stratification dimensions summarized in Table I: Azure resource type, complexity band, criticality tier, processing style, development mode, and business domain. Values are as recorded in the organization’s interface catalogue and the study’s data-collection records; interface names are omitted for confidentiality.

TABLE IV
PER-INTERFACE CHARACTERISTICS OF THE FIFTEEN STUDY INTERFACES

| Interface ID | Azure Resource Type | Complexity | Criticality | Processing Style | Dev. Mode | Business Domain |
|---|---|---|---|---|---|---|
| EAII-215 | Data Factory + SQL | High | Silver | Transformation | Greenfield | B2B Trading |
| EAII-216 | Data Factory + SQL | High | Silver | Transformation | Greenfield | B2B Trading |
| EAII-249 | Logic App + MFT | Very High | Gold | Transformation | Greenfield | Supply Chain (NZ) |
| EAII-258 | SQL | High | Gold | Mapping-only | Brownfield | Retail |
| EAII-259 | Function App + SQL | Very High | Gold | Transformation | Greenfield | Retail |
| EAII-260 | Function App | High | Gold | Mapping-only | Brownfield | Retail |
| EAII-265 | Logic App | High | Gold | Transformation | Greenfield | Corporate |
| EAII-266 | Logic App | High | Silver | Transformation | Greenfield | B2B Trading |
| EAII-270 | Function App | Medium | Gold | Transformation | Greenfield | Retail |
| EAII-318 | Logic App | Medium | Gold | Orchestration | Greenfield | Supply Chain (NZ) |
| EAII-324 | Logic App | Low | Gold | Pass-through | Greenfield | B2B Trading |
| EAII-367 | Function App | Very High | Gold | Transformation | Greenfield | Corporate / Enabling Services[a] |
| EAII-436 | Logic App + Function App + Blob | Very High | Silver | Transformation | Greenfield | Marketing Analytics |
| EAII-505 | Logic App + MFT | Medium | Gold | Transformation | Greenfield | Logistics |
| EAII-510 | Data Factory + Blob | Medium | Bronze | Transformation | Greenfield | Marketing Analytics |

[a] Provides shared enabling services to the corporate domain; counted under both business domains in Table I.

## APPENDIX B: COT PROMPTING EXAMPLE SEQUENCE

The following four prompts are reproduced from the team's live proof-of-concept demonstration of the framework on interface EAII-270 ("Client Identity"). They illustrate the staged chain-of-thought progression that Section IV formalizes as the six-stage pipeline: exemplar analysis, content extraction and template mapping, scoped content generation, and diagram generation. The prompts are shown verbatim; only list formatting has been normalized for presentation.

**Prompt 1: Exemplar and template analysis**

```
You are assisting in the generation of a Technical
Specification Document (TSD) for an EAII interface.

You are provided with the following two documents:
1. Example TSD: TSD Sales Extract (completed reference TSD)
2. Example FSD: FSD Sales Extract (corresponding functional
specification)

PHASE 1 OBJECTIVE:
Do NOT generate any TSD content yet.

Your task is to carefully review and analyze all TSD
documents to understand:
• How the TSD is structured and formatted based on the
template
• How information from an FSD is translated into a
completed TSD
• The level of technical depth, tone, and detail expected
in a production TSD
• How business and functional requirements are mapped to
technical sections
• Common patterns in scope definition, assumptions,
constraints, and terminology

After reviewing, provide:
1. A summary of how the example TSD was derived from its
corresponding FSD
2. Key observations about structure, formatting, and
section depth
3. Any rules or patterns that should be followed when
generating a new TSD
4. Confirmation when you are ready to proceed to content
extraction

Do not proceed to any generation or extraction until this
analysis is complete.
```

**Prompt 2: FSD content extraction and template mapping**

```
PHASE 2 OBJECTIVE:
Extract and organize information, but do NOT generate final
TSD content yet.

Using:
• Functional Specification: EAII-270 – Client Identity – BP
GUI to APIM
• Standard TSD Template (enterprise-approved template)

Perform the following tasks:

PART A — FSD CONTENT EXTRACTION:
From the EAII-270 Functional Specification, extract and
summarize:
• Business purpose and background
• Source and target systems
• Supported operations and capabilities
• Scope (what is included and excluded)
• Assumptions, constraints, and dependencies
• Any architectural or integration-related details relevant
to a TSD
Organize this information into clearly labeled bullet
points.

PART B — TEMPLATE INTERPRETATION:
Review the Standard TSD Template and explain:
• The intent of Sections 1 and 2
• What type of content belongs in each subsection
• How extracted FSD information should map into these
sections

Output:
• A structured summary of extracted FSD content
• A clear mapping explanation of how this content will be
used in Sections 1 and 2

Do not generate any final TSD prose yet.
Wait for confirmation before proceeding.
```

**Prompt 3: Scoped TSD content generation (Sections 1–2)**

```
PHASE 3 OBJECTIVE: Generate production-quality TSD content
using only previously analyzed and extracted information.

Using:
• The Standard TSD Template
• The extracted EAII-270 content from Phase 2

Generate content ONLY for the following sections of the
Technical Specification Document:

SECTION 1 — Document Overview
• Background
• Document Purpose
• Intended Audience (if applicable)

SECTION 2 — Scope
• In Scope
• Out of Scope
• Assumptions and Constraints

Guidelines:
• Follow enterprise documentation tone and style
• Ensure content is precise, technical, and
implementation-aware
• Do not include architecture diagrams or sequence diagrams
yet
• Do not generate content for any other sections
• Ensure alignment with EAII integration standards

The output should read as a clean, review-ready draft
suitable for engineering teams.
```

**Prompt 4: Integration sequence diagram**

```
PHASE 4 OBJECTIVE: Create a high-level integration sequence
diagram for EAII-270.

Based on your understanding of the interface, generate a
sequence diagram that illustrates:
• BP GUI as the initiating system
• APIM as the API gateway
• Downstream Azure components involved
• Request, processing, and response flow at a high level

Use a clear, readable diagram format (PlantUML or
equivalent text-based representation).

Guidelines:
• Focus on logical flow, not low-level implementation
• Ensure participant naming is clear and
enterprise-appropriate
• Keep the diagram aligned with EAII integration patterns

This phase is independent of document generation and
focuses only on visualizing the integration flow.
```

## ACKNOWLEDGMENT

The authors thank the client organization's program sponsors and delivery management for supporting the study and for reviewing and approving its artifacts, and the joint client–consultant delivery team for the collaboration that made the measurement program possible.

**Disclosure of AI use.** Generative AI was used in one capacity. GitHub Copilot (with GPT-4.1/5 and Claude Sonnet 4) was the object of study and produced the technical-specification drafts analyzed herein.

**Data availability.** The generated TSDs, the Prompt-as-a-Code repository, and the time-tracking workbook reside in the partner organizations' internal systems (Azure DevOps, SharePoint, Confluence) and cannot be shared publicly. Anonymized excerpts are available from the authors on reasonable request, subject to organizational approval.

## REFERENCES

1. M. Chen et al., “Evaluating large language models trained on code,” arXiv:2107.03374, 2021.
2. A. Ziegler et al., “Productivity assessment of neural code completion,” in Proc. 6th ACM SIGPLAN Int. Symp. Machine Programming (MAPS), 2022, pp. 21–29.
3. S. Peng, E. Kalliamvakou, P. Cihon, and M. Demirer, “The impact of AI on developer productivity: Evidence from GitHub Copilot,” arXiv:2302.06590, 2023.
4. X. Hou et al., “Large language models for software engineering: A systematic literature review,” ACM Trans. Softw. Eng. Methodol., vol. 33, no. 8, 2024.
5. J. Wei et al., “Chain-of-thought prompting elicits reasoning in large language models,” in Adv. Neural Inf. Process. Syst. (NeurIPS), 2022.
6. A. Vaswani et al., “Attention is all you need,” in Adv. Neural Inf. Process. Syst. (NeurIPS), 2017.
7. T. B. Brown et al., “Language models are few-shot learners,” in Adv. Neural Inf. Process. Syst. (NeurIPS), 2020.
8. A. Svyatkovskiy, S. K. Deng, S. Fu, and N. Sundaresan, “IntelliCode Compose: Code generation using transformer,” in Proc. ESEC/FSE, 2020, pp. 1433–1443.
9. P. Vaithilingam, T. Zhang, and E. L. Glassman, “Expectation vs. experience: Evaluating the usability of code generation tools powered by large language models,” in Proc. CHI Conf. Human Factors in Computing Systems, Extended Abstracts, 2022.
10. S. Barke, M. B. James, and N. Polikarpova, “Grounded Copilot: How programmers interact with code-generating models,” Proc. ACM Program. Lang., vol. 7, no. OOPSLA1, 2023.
11. S. I. Ross, F. Martinez, S. Houde, M. Muller, and J. D. Weisz, “The programmer’s assistant: Conversational interaction with a large language model for software development,” in Proc. 28th Int. Conf. Intelligent User Interfaces (IUI), 2023, pp. 491–514.
12. A. Fan et al., “Large language models for software engineering: Survey and open problems,” in Proc. IEEE/ACM ICSE – Future of Software Engineering (ICSE-FoSE), 2023, pp. 31–53.
13. X. Hu, G. Li, X. Xia, D. Lo, and Z. Jin, “Deep code comment generation,” in Proc. 26th IEEE/ACM Int. Conf. Program Comprehension (ICPC), 2018, pp. 200–210.
14. A. LeClair, S. Jiang, and C. McMillan, “A neural model for generating natural language summaries of program subroutines,” in Proc. 41st IEEE/ACM Int. Conf. Software Engineering (ICSE), 2019, pp. 795–806.
15. J. Y. Khan and G. Uddin, “Automatic code documentation generation using GPT-3,” in Proc. 37th IEEE/ACM Int. Conf. Automated Software Engineering (ASE), 2022.
16. D. Nam, A. Macvean, V. Hellendoorn, B. Vasilescu, and B. Myers, “Using an LLM to help with code understanding,” in Proc. 46th IEEE/ACM Int. Conf. Software Engineering (ICSE), 2024.
17. T. Kojima, S. S. Gu, M. Reid, Y. Matsuo, and Y. Iwasawa, “Large language models are zero-shot reasoners,” in Adv. Neural Inf. Process. Syst. (NeurIPS), 2022.
18. P. Liu et al., “Pre-train, prompt, and predict: A systematic survey of prompting methods in natural language processing,” ACM Comput. Surv., vol. 55, no. 9, 2023.
19. J. White et al., “A prompt pattern catalog to enhance prompt engineering with ChatGPT,” arXiv:2302.11382, 2023.
20. C. Ebert and P. Louridas, “Generative AI for software practitioners,” IEEE Software, vol. 40, no. 4, pp. 30–38, 2023.
21. P. Runeson and M. Höst, “Guidelines for conducting and reporting case study research in software engineering,” Empirical Software Engineering, vol. 14, no. 2, pp. 131–164, 2009.
22. GitHub, Inc., “GitHub Copilot documentation.” [Online]. Available: https://docs.github.com/copilot (accessed Aug. 2026).
23. OpenAI, “GPT-4 technical report,” arXiv:2303.08774, 2023.
24. Anthropic, “Introducing Claude 4,” May 2025. [Online]. Available: https://www.anthropic.com/news/claude-4
25. T. Wu, M. Terry, and C. J. Cai, “AI Chains: Transparent and controllable human-AI interaction by chaining large language model prompts,” in Proc. CHI Conf. Human Factors in Computing Systems (CHI), 2022, pp. 1–22.
26. V. Gadesha and E. Kavlakoglu, “What is prompt chaining?” IBM, Apr. 2024. [Online]. Available: https://www.ibm.com/think/topics/prompt-chaining (accessed Aug. 2026).
27. B. A. Kitchenham et al., “Preliminary guidelines for empirical research in software engineering,” *IEEE Trans. Softw. Eng.*, vol. 28, no. 8, pp. 721–734, Aug. 2002.

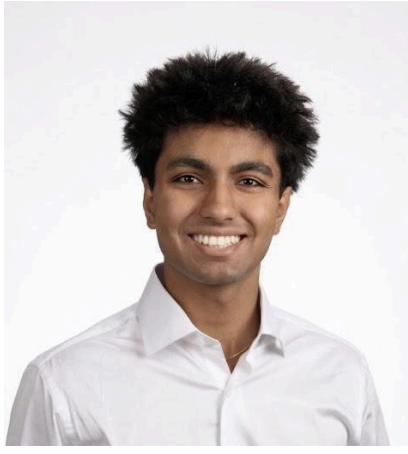

**ARYAN DHINGRA** is a software engineer and AI researcher whose work sits at the intersection of technology, enterprise systems, and entrepreneurial innovation. He is a student at Bergen County Academies — Academy for Engineering Design Technology (AEDT), Hackensack, NJ. A USACO Gold Division qualifier and AWS Certified Cloud Practitioner; he has built professional depth through progressive roles. In 2024, he was with Khoj Information Technology, working in Azure cloud computing. In 2025, he was with Metcash, where he engineered prompt-driven systems automating enterprise documentation workflows. He is currently with Morgan Stanley, New York, NY, in enterprise computing and database engineering. He completed the Wharton Global Youth Essentials of Entrepreneurship program, deepening his business-technology perspective.

He is the sole inventor of AI Caretaker, a voice-activated in-home safety system for dementia patients (USPTO patent pending). His research centers on prompt engineering, large language models, and scalable AI deployment in healthcare and enterprise systems.

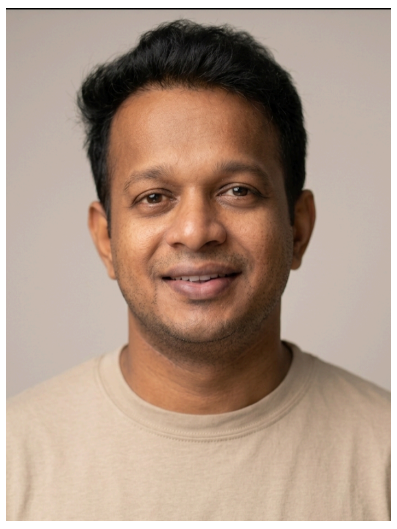

**GOKUL MENON** is a Microsoft certified Senior Cloud & AI Engineer with over 8 years of industry experience building software, cloud & AI solutions for global clients. He holds a master's degree in computer science from Newcastle University, UK, and multiple Microsoft certifications spanning Azure and applied AI. At Khoj Information Technology, USA, he has architected, developed, and deployed AI solutions including an autonomous error-classification and root-cause-analysis agent that combines retrieval-based classification with GPT-5-driven diagnostics for an enterprise monitoring platform, and a GitHub Copilot-driven development workflow that cut engineering time by 60% across multiple teams. His work also spans Azure services for automating cloud integration across enterprise systems. Earlier in his career, he specialized in secure systems engineering, contributing to C++-based cybersecurity auditing software supporting compliance with standards such as DISA STIG, CIS, and ISO/IEC 27001, alongside secure coding and DevSecOps practices. He also worked extensively in game development, building and optimizing gameplay systems in Unreal Engine for AAA titles across console and PC platforms, and led quality-assurance and test-automation efforts, including automated testing pipelines for large-scale game releases, before transitioning into cloud and AI engineering. His research interests include cloud-native architecture, applied artificial intelligence for enterprise systems, and automated software engineering

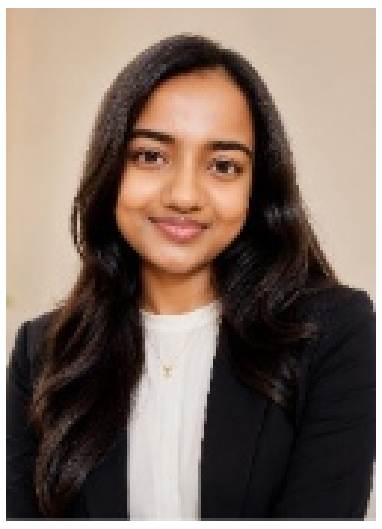

**ANUSHA ANTONY** is a Project Management Analyst at Khoj Information Technology with over four years of experience coordinating cross-functional delivery teams and optimizing enterprise project workflows. She holds a Bachelor of Computer Applications (BCA) from Maharshi Dayanand University, Rohtak, India (2021). She has managed large-scale initiatives spanning enterprise system builds, application security remediation programs, and AI-assisted development proofs-of-concept, including the GitHub Copilot–assisted development work described in this article, and is highly skilled in using Azure DevOps Boards for effective project planning, tracking, and execution. Her interests include AI-assisted software delivery, the adoption of generative AI tools in enterprise engineering programs, and project management practices for large-scale technology transformations.